\documentstyle[12pt,aasms4]{article}

\slugcomment{to appear in proceedings of the STScI Symposium ``Astrophysics of Life'' 6-9 May, 2002}

\lefthead{de la Fuente et al.}
\righthead{}

\begin{document}
\title{Protoplanetary Disks in the Orion Nebula: An H$\alpha$ Fabry-Perot study and Astrobiological Aspects}
\author{Eduardo de la Fuente, Margarita Rosado, Lorena Arias, Patricia Ambrocio-Cruz}
\affil{Instituto de Astronom\'\i a, UNAM, Apdo. Postal 70-264, C.U., M\'exico, D.F. \\
{\tt edfuente, margarit,lorena,patricia@astroscu.unam.mx}}

\author{and\\
  Henry B. Throop}
\affil{University of Colorado, Boulder, CO 80309-0389, USA. \\
{\tt throop@ciclops.swri.edu}}


\begin{abstract}
{\bf In this paper, we present a briefly overview of the protoplanetary disks in the Orion Nebula, incluiding some astrobiological aspects and an H$\alpha$ Fabry-Perot study of 16 of them. We found that Fabry-Perot interferometry constitutes an effective technique for the detection of proplyds. We also report heliocentric systemic velocities for the proplyds 82-336, 158-323, 158-326, 159-350, 161-314, 161-324, 163-317, 166-316, 167-317, 168-326, 170-337, 176-325, 177-341, 180-331, 197-427 and 244-440. The velocities were measured between 22-38 km s$^{-1}$.}

\end{abstract}


\section{INTRODUCTION}

The proplyds (PROtoPLanetarY DiskS) are photoevaporating circumstellar disks around young stellar objects (YSOs). They are composed of: 1.- a circumstellar disk around the YSOs, 2.- a head with a bright cusp, 3.- stand-off shocks and 4.- a tails. Their shapes are created when the UV photoionizing flow from a hot (O spectral type) external star interacts with a photoevaporated flow from the circumstellar disk: the photoionizing flux (2000 $>$ $\lambda$ $>$ 912~ \AA) infringes on the disk producing a slow ($\sim$ 3  km s$^{-1}$) photodissociated flow.  The flow passes through a weak D-type ionization front (IF) and is accelerated to transonic velocities ($\sim$ 10 km s$^{-1}$) producing a weak shock that lies just inside the IF, generating a slight density and column density enhancement that is seen in silhouette (head). The electron density (n$_e$) at the IF has values of $\sim$ 10$^{6}$-10$^{5}$ cm$^{-3}$. The greater values correspond to the inner regions. The 10$^{5}$ cm$^{-3}$ density is reached just outside of the IF. The n$_e$ decreases with the distance from the YSO as r$^{-2}$ due to the quasi-spherical geometry of the photoevaporated flow. The bright proplyd cusp corresponds to the photoevaporated flow from the head of the proplyd driven by the ionizing photons from the hot external star. Therefore, the head is composed of check ionization fronts and photoablated material from the circumstellar disk. The photoablated material expands outwards from this IF. When its ram pressure  is balanced with the ram pressure of the photoionizing flow or stellar wind of the hot star, an arc-shaped stand-off shock forms, which is visible in H$\alpha$ and in [OIII] lines. In these arcs, n$_e$$\sim$10$^{4}$ cm$^{-3}$. If the stellar infringing outflow is subsonic the H$\alpha$+[OIII] arcs are not formed. Finally, the tails are generated by the photoevaporated flow driven by the diffuse UV photon radiation from the nebula. These tails have n$_e~\sim$10$^{5}$ cm$^{-3}$ near to the head and $\sim$10$^{4}$ cm$^{-3}$ at the tip. Recently, microjets have been observed in some proplyds [22, 3]. Detailed information about this scheme for proplyd formation is presented by [1], see also Figure 5 of [14].
More detailed,  the actual proplyd formation schemes  propose that the proplyds are formed and detected due to the photoevaporation of the protoplanetary disk by the intense UV radiation of an external star (in this case $\Theta^{1}$ Orionis C and, for some other proplyds, together with $\Theta^{2}$ Orionis A). It is assumed that the protoplanetary disk is surrounded by a neutral envelope probably formed by a slow photodissociated wind from the disk. It is also assumed that an IF, forms in this envelope and that the newly ionized gas flows away from the proplyd with an initial velocity v$_0$. Pressure gradients in the ionized gas  accelerate the flow away from the IF. The amount of the acceleration depends on the type of the IF, being maximal when the flow leaving the IF surface is sonic (D-critical IF). The density in the ionized photoevaporated flow is highest at the point on the IF closest to the ionizing star, n$_0$, and decreases towards the sides. Assuming the flow to be isothermal and r$_0$ the radius of the IF to the point of maximum density, then the photoevaporated flow initial velocity, v$_0$, is related to the disk's photoevaporation mass-loss rate via the relation:
$\dot{M}$ = 4$\pi$ r$_0$$^{2}$ v$_0$ n$_0$ m$_I$ (where m$_I$ = 1.35 m$_H$) or [$\dot{M}$/10$^{-7}$ M$_{sun}$~ yr$^{-1}$] = 0.44 [r$_0$/10$^{15}$ cm]$^{2}$ [v$_0$/km s$^{-1}$] [n$_0$/cm$^{-3}$] and the life time, $\tau$, of the protoplanetary disk is given by  $\tau$ $\sim$ M$_{disk}$/$\dot{M}$ [14, 16]. Thus, to obtain the radial velocity profiles of proplyds is important because it allows us to constrain the photoevaporated flow velocity. This quantity gives us an estimate of the mass-loss rates of the protoplanetary disks and of the life time of those disks. The scheme described above is also presented by [13, 11, 14]. and are based on the photoevaporating flow models developed  by [9, 4, 13, 16, 34].

The proplyds were discovered by Laques and Vidal in 1979 [17]. The term proplyd  was coined by C.R. O'Dell to name the 6 nebulosities called LV objects identified by these authors. The proplyds were identified as young stars with circumstellar clouds photoionized from the exterior by $\Theta^{1}$ Orionis C by means of VLA observations [6, 10]. They were imaged in 1993 with HST to show, for the first time, their morphology: silhouettes, bow shocks and even circumstellar disks [24, 25]. Indeed they were well understood to be a compact disks around YSOs with these excellent imagery. Recently the proplyds are designated acoording the notation of O'Dell \& Wen [25]. This designation system is based on the proplyds ecuatorial (J2000) coordinates (1" in declination and 1.5" in Right Ascension). By example, the proplyd 231-739 has coordinates ($\alpha$=5$^{h}$ 35$^{m}$ 23.123$^{s}$, $\delta$=-5$^{\circ}$ 27$^{m}$ 38.9$^{s}$). 

The proplyds have $\dot{M}$ of 10$^{-6}$ - 10$^{-7}$ M$_{sun}$  yr$^{-1}$ [6, 14], radial velocities of 24 to 30 km s$^{-1}$ [14], diameters of 10$^{-4}$ - 10$^{-3}$ pc [28] and disk masses of 0.005 - 0.02 M$_{sun}$ [2, 16]. These values are estimated from extinction measures of the silhouette disks on the background HII region emission [36] or from CO milimetric observations of the thermal dust emission from their disks [2]. These latter observations give evidence of accretion disks surrounding young stellar objects ([34] and references therein). In the Orion Nebula (M 42, NGC 1976) there are $\sim$ 150 proplyds [27], 21 present microjets (see Figure 7a of [3]) and 15 appear in pure silhoutte; YSOs and circumstellar disk (see Figure 7a of [3]). A recent review about the Orion nebula is presented by O'Dell [29].

\section{ASTROBIOLOGICAL ASPECTS}

Given that in the protoplanetary disks, planet formation can occur and, as a consequence, it is possible that the formation of life can also occur there, the interest in astrobiological studies of the proplyds stems from the fact that  proplyds unveil the otherwise ellusive protoplanetary disks. In this sense, proplyds are an important laboratory for determining, at high spatial resolution, several properties of the protoplanetary disks such as: masses, velocities and extensions.  The proplyds are also useful in statistical studies on these issues. Furthermore, the protoplanetary disks revealed by the proplyds are subject to the strong UV flux of the exterior star that makes them visible. Thus, the photoevaporation of the protoplanetary disk of the proplyd could inhibit planet formation or change its conditions. Recent studies [2, 36] have shown that planet formation in irradiated protoplanetary disks as proplyds can only occur in special situations and conditions: disk mass $\geq$ 0.13 M$_{sun}$ and dust (silicates+ices) particle radius $\geq$ 5 $\mu$m.

Furthermore, Throop et al. [36] through H$\alpha$ and Pa$\alpha$ HST images, obtained an extintion curve of the biggest (in disk) proplyd 114-426 finding that a theoretical extintion curve dominated by dust grains $\geq$5 $\mu$m fixed well the observations. They also present 1.3mm OVRO observations suggesting that the dust growth reaches radius values of few mm, and developed a numerical model to explain the behaivor and evolution of dust grains in proplyds irradiated externally including photo-destruction process. With this, they have shown that after 10$^5$ yr, small grains are entrained in the photoevaporative flow resulting in disks with maximum sizes of 40 AU. In less than 10$^5$ yr, in disk masses as large as 0.2 M$_{sun}$, the grain growth reaches a radius of meters at 10 AU of the YSO and of 1 mm at 500 UA from it. These sizes allow the dust to resist the photoevaporation process.  By 10$^6$ yr,  nearly all ice and gas are removed by photosputtering inhibiting the Kuiper belts formation and leaving only the possibility of rocky planet formation. Therefore, it is possible that in the environments of star forming places like Orion, Jupiter-like planets are not formed in the standard way because it would require 10$^6$ to 10$^7$ yr ([33] and references therein). Such Jovian planets could be present if they form on 10$^3$ yr time-scale [5] in disk masses $\geq$ 0.13 M$_{sun}$. Since most of the stars seem to be formed in large and dense clusters such as Orion, this leads to the conclusion that planet formation models should be revised in order to include the destructive effects of the UV flux of newly formed massive stars in star forming environments. The possibilities for planetary system to form in the Orion Nebula are described by Throop et al. [37].

In summary, in enviroments with external hot stars like the proplyds in the Orion Nebula: 1.- The formation of Jovian planets and kuiper belts objects is very dificult, terrestrial planets remain unaffected and posible. 2.- The solar systems formed here must be different from our solar system, in fact, the standar solar nebula models can not be applied here. The important parameter to determinate the planet formation time scales is $\dot{M}$. Indeed, the typical values suggest life time for the protoplanetary disks of 10$^5$ yrs. This last involve that in many of this proplyds, the planet formation can not occurr. Maybe the 15 proplyds suggested by Bally and co-workers (see [3] and Bally in this symposium) are factible places for planet formation.     

\section{FABRY-PEROT STUDY}

With Fabry-Perot (FP) interferometry and long-slit spectrographs, it is possible to study the kinematics of every astronomical object. Both techniques are useful and complementaries. The fundamental difference between both instruments is that with a FP is possible to study the whole region because the instrument cover all the field although the information is concentrated in a small part of the electromagnetic spectrum. In the spectrograph occurs the opposite: it is possible to obtain a spectrum but in a very limited zone of the object. Due to the fact that Orion is a nebula rich in interstellar phenomena and that it displays a gradient in the radial velocity across its area, we can, with FP techniques, identify the different velocity components of a radial velocity profile, in this case, the HII region velocity component and the proplyd velocity component. Spectroscopic studies in [OIII] of the proplyds reported in [17] were made by [21, 19, 20, 22, 12] using slit spectrometers. Keck high resolution slit spectroscopy of proplyds at several ions including H$\alpha$, [SII] and [OIII]  is presented by [14]. On the other hand, [26] present a Fabry-Perot kinematical study at [OIII](5007\AA) and [SII](6731\AA) of the inner region of the Orion Nebula, discovering new high velocity features and studying the redshifted and blueshifted emission of some proplyds. With these spectroscopical studies it is possible to determine the flow velocity, and from the [SII] lines and/or H$\alpha$ emission measure, the density of the IF [1]. Knowing the systemic velocities and the radial velocity profiles of the proplyds we can compare the observations with the existing theoretical models in order to present arguments in favor of the existence of accretion disks in them, and know the behavior between the interaction of the photoevaporated material and the photoionizing winds. Also, from the determination of the velocity of the gas we can derive the $\dot{M}$ in combination with observations of the surface brightness in H$\alpha$ [28].

This paper and [7] complements the previous works [31, 32], where a study of the kinematics of the gas, Herbig-Haro (HH) objects, and jets at large scales in the Orion Nebula are presented. These works between other things, find that HH 202, HH 203 and HH 204 are part of a large bipolar outflow centered near the E-W jet discovered by [26]. This last is suggested because HH 202 form part of a big blue-shifted lobe, whereas HH 203-204 form part of a other big lobe that present red-shifted emission in our data cubes (e.g. see last panels on Figures 5 and 6 of [31]). Furthermore, recent proper motion study of these HH objects suggest that they emerge of the same source [8]. 

\subsection{Observations, technique and results}

Fabry-Perot data cubes were obtained from November 30 to December 5, 1996 with the ``PUMA'' scanning Fabry-Perot spectrograph [30] at the f/7.5 Cassegrain focus of the 2.1m telescope of the Observatorio Astron\'omico Nacional at San Pedro M\'artir B.C., M\'exico. See [31] for further details. The data reduction was made using the software CIGALE [18]. The results are presented in detail by [7].

\begin{table*}[ht] \scriptsize
 \centering
 \begin{minipage}{140mm}
  \caption{Positions and velocity extent of the identifed proplyds} 
  \begin{tabular}{@{}cccccccccc@{}}
Proplyd Name $^a$   & x$^b$ & y$^b$ & $V_{\rm HII~region}$ & $V_{proplyd}$ & $V_{bs}$ &  FWHM proplyd & FWHM corrected\\  & & & (km s$^{-1}$)  &   (km s$^{-1}$)  &   (km s$^{-1}$)  &  (km s$^{-1}$)  & (km s$^{-1}$)  \\ [10pt]
\hline
82-336  & 193 & 251 & 13 & 33 & 37 & 18.92 & 6.5 & \\
158-323 (LV 5) & 254 & 228 &  9 & 34 & 32 & 18.92 & 6.5 & \\
158-326 (LV 6) & 255 & 235 & 11 & 33 & - & 28.38 & 20.14  \\
159-350 & 251 & 275 &  9 & 37 & 33 & 18.92 & 6.5 & \\
161-314 & 246 & 211 &  9 & 27 & 14 & 18.92 & 6.5 & \\
161-324 & 248 & 231 &  9 & 34 & - & 9.46 & 17.62 \\
163-317 (LV 3) & 242 & 218 & 12 & 24 & - & 18.92 & 6.5 & \\
166-316 & 236 & 212 & 11 & 27 & - & 18.92 & 6.5 & \\
167-317 (LV 2) & 231 & 218 & 12 & 28 & - & 18.92 & 6.5 & \\
168-326 (LV 1) & 228 & 235 &  9 & 36 & - & 28.38 & 20.14 & \\
170-337 & 224 & 253 &  13 & 22 & - & 28.38 & 20.14 & \\
176-325 & 210 & 231 & 9 & 36 & 32 & 18.92 & 6.5 & \\
177-341 & 205 & 259 & 13 & 32 & 28 & 37.84 & 32.12 & \\ 
180-331 & 197 & 241 & 12 & 38 & 24 & 18.92 & 6.5 & \\
197-427 & 202 & 295 &  5 & 30 & - & 18.92 & 6.5 & \\
244-440$^c$ & 34  & 361 & 19$^c$ & 23$^c$ & - & 47.30$^c$ & 42.86$^c$ & \\
\hline
\end{tabular}
$^a$According with the notation of O'Dell \& Wen [25] \\
$^b$Pixel position in our data cubes \\
$^c$See the text \\
\end{minipage}
\label{sampletabel}
\end{table*}

\subsubsection{Unsharp-masking technique and proplyds identification}

We were able to identify the proplyds: 82-336, 158-323, 158-326, 159-350, 163-317, 167-317, 170-337, 177-341 and 244-440 from our raw FP data cubes (see Figure 1 of [31, 32]. These proplyds show a conspicuous appearance in some of the velocity maps: they are detected  as bright, point-like nebulosities. In order to improve the detection of proplyds from our velocity maps we applied an ``unsharp-masking'' process in order to isolate the  small diameter emission structures from the bright diffuse HII region (as it will be described below). This technique is also useful for detecting extended features embedded in bright emission (e.g. [23]). It consist in creating an out-of-focus image which is used as a spatial frecuency filter, to enhance rendition of any fine details, while at the same time reducing gross density variations on the resulting filtered image. Indeed, this process improves the detection of proplyds. The proplyds: 161-314, 161-324, 166-316, 168-326, 176-325, 180-331 and 197-427 became noticeable only after applying the unsharp-masking  process while the proplyds already detected in the raw data cubes were more easily distinguished.

Figure 1 shows the resulting FP velocity map at V$_{helio}$ = - 127 km s$^{-1}$ (presented in [31, 32]) after applying the unsharp-masking  process. In this map we have marked the position of the identified proplyds that, as seen in this figure, are quite conspicuous. The unsharp-masking  process has been carried out in the following way: first, we have done a spatial, Gaussian smoothing (with $\sigma$ = 3 pixels or 1.77 $\arcsec$) that smoothes small diameter features (such as the proplyds and thin filaments). Then, we subtracted the smothed velocity maps from the original velocity maps. The result is that the small diameter features are sorted out whereas the diffuse, extended emission (such as the foreground HII region) is subtracted. In that way we were able to identify the proplyds reported in [28]. The pixel coordinates of the identified proplyds are quoted in Table 1 where the proplyds' names according to O'Dell \& Wen are given in column 1, whereas the proplyds pixel coordinates, x and y, are given in columns 2 and 3 respectly.

\subsubsection{Proplyds' radial velocity profiles and profile decomposition}

Once we have obtained the pixel coordinates of the identified proplyds, we proceeded to extract radial velocity profiles integrated over boxes of 2 $\times$ 2 pixels centered at the pixel positions reported in Table 1. For the obtainment of the radial velocity profiles we use only the original H$\alpha$ data cubes (i.e., before applying the unsharp masking process). The radial velocity profiles obtained in this way are contaminated by the bright HII region velocity component (proplyd+HII region profiles). We preferred to subtract the HII region contribution point by point taking advantage of the fact that our FP data cubes give also the velocity profiles of neighboring regions outside the proplyd position. In this way, in order to know the contribution of the HII region emission, we extracted radial velocity profiles of several zones close to the proplyd location. This allows us to identify the average intensity and velocity of the HII region velocity component (HII region velocity profile). Finally, using this information, we fitted in the proplyd+HII region profile, the HII region velocity profile and a remaining velocity component that we identify as the proplyd velocity profile. The main problem and difficulty in the kinematical  studies of the proplyds resides in separating the velocity component of the proplyd from those of the HII region because the intensity of the HII region with respect to the proplyd is higher. Furthermore, both the proplyd and the HII region can have similar or same radial velocity making the task quite difficult. However, the use of FP techniques that cover all the field around the proplyd, allow us to get the radial velocity profile of the HII region quite accurately. Consequently, we were able to obtain the proplyd profiles with an accuracy of $\pm$ 5  km s$^{-1}$ in the peak velocities and $\pm$ 10  km s$^{-1}$ in the FWHM's.

Figures 2, and 3 show the radial velocity profiles obtained at the identified proplyds positions, as described above. The radial velocity profiles shown in these figures correspond to the proplyd+HII region profiles. The velocity components of  the HII region and those assumed to be of the proplyds are also shown in this figure. Table 1 also quotes the heliocentric peak radial velocities of the contaminating HII region and the proplyds, as well as the FWHM of the proplyds (columns 4, 5 and 7 respectively). The HII and proplyd velocity components are in agreement with typical values cited in the literature (e.g. [28, 14]). As it is discussed in the study by [26], many proplyds present redshifted high velocity flows. For 159-359, 82-336, 180-331, 175-325 and 161-314 we also found weak blueshifted components (column 6 in Table 1) probably related to the HII region emission (Figures 2 and 3). About these components, a better spatial resolution studies are needed to clarify their origin. Unfortunately, we can only compare our results with the velocity profiles obtained from spectroscopic H$\alpha$ data of 3 proplyds (170-337, 177-341 and 244-440) of [14]. We find a good agreement with 170-337 for the proplyd peak velocity (22 vs 20 km s$^{-1}$) while for  244-440 and 177-341  we  find that the proplyd peak velocity is different from the one reported by [14] of 23 vs 10 km s$^{-1}$ and 32 vs 22  km s$^{-1}$, respectively, in spite that those proplyds appear bright in our data cubes. The values reported in Table 1 for 244-440 cover all its area. Indeed, a more detailed inspection (see [7]) suggest that the center (1$\times$1 pixels) of the proplyd have V$_{helio}$ = 16 km s$^{-1}$, whereas the regions around it have values of 18-23 km s$^{-1}$. The detection of proplyds in our [NII] data cubes is poor. Indeed, we only identify two of them: 170-337 and 244-440. Howewer, they have radial velocities of 29 and 12 km s$^{-1}$ that are in agreement with the results reported by [14] of 33 and 10 km s$^{-1}$ respective!
ly. It is very important to carry out 3-dimensional spectroscopy especially in H$\alpha$, [OI] and [NII], with better spectral resolution but covering the entire field of view, in order to verify whether there is agreement or not with the results reported here.
\\

Our results do not allow us to study the proplyd profile, including their diferent components and even their forms (not necessarily Gaussian). We can only find other velocity components different from the HII region, proposing that these velocities are related with the proplyds. 
Table 1 lists the proplyd peak velocities as well as the velocities of the contaminating HII regions obtained from our FP velocity profiles. As one can see from this table, all the proplyd peak velocities are redshifted relative to the HII region indicating that all the proplyds studied in this work are located in the same gaseous sheet behind the HII region. In addition, Table 1 reports the proplyd velocity widths uncorrected and once corrected from the instrumental and thermal widths. These widths have values of 10 and 20 km s$^{-1}$ respectly, similar to the presented by [14]. This table also quotes the H$\alpha$ velocity widths (FWHM) obtained in this work, corrected for the instrumental function and thermal broadening (column 8). An inspection of the values of this quantity for all the proplyds we study here shows that the velocity widths are either smaller or comparable to the ones reported by [14], and that the H$\alpha$ velocity widths do not seem to vary as a function of distance to the exterior ionizing source.  Consequently, we expect that the initial flow velocity is comparable to the one found by these authors, i.e. v$_0$=13 km s$^{-1}$. With this value and taking n$_0$, r$_0$ from the literature [18] (as is discussed and presented in de la Fuente et al. [7]), we found $\dot{M}$ values between 10$^{-6}$ to 10$^{-7}$ M$_{sun}$ yr $^{-1}$ and $\tau$ $\sim$ 10$^{4}$-10$^{5}$ yr. By example, the proplyd 161-314 have $\dot{M}$ $\sim$ 2$\times$10$^{-7}$ M$_{sun}$ yr $^{-1}$ and $\tau$ $\sim$ 5$\times$10$^{5}$ yrs. Thus, our kinematic work confirms the conclusions on time scales for disk destruction of 10$^{5}$ yr [13]. Futhermore, our $\tau$ values suggest that in our proplyds sample, the planet formation could be difficult and indeed, implies the necessity to revise the models of planet formation as discussed by [36, 37]. It should be quite valuable to obtain FP observations at [OI](6300 \AA) in order to derive with  better accuracy, the $\dot{M}$ of each proplyd. Futhermore there is some evidence th!
at v$_0$ is traced by this ion [14]. 
\\

MR wishes to acknowledge the financial support from DGAPA-UNAM  via the grant IN104696. E de la F wishes to acknowledge financial support from CONACYT-M\'exico grant 27550-A. It is a great pleasure to thank John Bally for many informative comments and guidelines.

\clearpage

\vfill\eject

\figcaption{ Unsharp masking H$\alpha$ velocity map at heliocentric velocity = -127 km s$^{-1}$. The original image published by Rosado et al. [31, 32] is shown in the insert. The identified studied objects are marked.  
\label{fig1}}

\figcaption{Radial velocity profiles of the detected proplyds 82-336, 158-326, 159-350, 161-314, 161-324 and 163-317 (longest component). The middle size component stands for the HII region. The smallest one represents the suggested proplyd. Some profiles feature an extra  blueshifted component. The squares represent the fit of these components.
\label{fig2}}

\figcaption{Same as Figure 2 for proplyds 166-316, 167-317, 168-326, 170-337, 176-325, 180-331, 158-323, 177-341 and 197-427.
\label{fig3}}

\end{document}